# Optical frequency comb spectroscopy at 3–5.4 µm with a doubly resonant optical parametric oscillator

AMIR KHODABAKHSH[1], VENKATA RAMAIAH-BADARLA[1], LUCILE RUTKOWSKI[1], ALEXANDRA C. JOHANSSON[1], KEVIN F. LEE[2], JIE JIANG[2], CHRISTIAN MOHR[2], MARTIN E. FERMANN[2], AND ALEKSANDRA FOLTYNOWICZ[1*]

[1]Department of Physics, Umeå University, 901 87 Umeå, Sweden
[2]IMRA America, Inc., 1044 Woodridge Avenue, Ann Arbor, Michigan, 48105, USA
*Corresponding author: aleksandra.foltynowicz@umu.se

2016-03-31

**We present a versatile mid-infrared frequency comb spectroscopy system based on a doubly resonant optical parametric oscillator tunable in the 3-5.4 µm range and two detection methods, a Fourier transform spectrometer (FTS) and a Vernier spectrometer. Using the FTS with a multipass cell we measure high-precision broadband absorption spectra of CH$_4$ and NO at ~3.3 µm and ~5.2 µm, respectively, and of atmospheric species (CH$_4$, CO, CO$_2$ and H$_2$O) in air in the signal and idler wavelength range. The figure of merit of the system is on the order of 10$^{-8}$ cm$^{-1}$ Hz$^{-1/2}$ per spectral element, and multiline fitting yields minimum detectable concentrations of 10-20 ppb Hz$^{-1/2}$ for CH$_4$, NO and CO. For the first time in the mid-infrared, we perform continuous-filtering Vernier spectroscopy using a low finesse enhancement cavity, a grating and a single detector, and measure the absorption spectrum of CH$_4$ and H$_2$O in ambient air at ~3.3 µm. © 2016 Optical Society of America**

Optical frequency comb sources in the mid-infrared (MIR) wavelength range (3-12 µm) have large potential for molecular spectroscopy, since the fundamental absorption bands of many species lie in this fingerprint region [1, 2]. The maximum achievable wavelength of low repetition rate (<1 GHz) direct comb sources is still limited to <3 µm, and longer wavelengths (>3 µm) can only be reached through nonlinear frequency conversion. Sources based on difference frequency generation (DFG) offer wide wavelength coverage in the MIR, but suffer from poor conversion efficiency [3, 4]. Higher average output power is provided by synchronously-pumped optical parametric oscillators (OPOs) [5, 6]. In particular, OPOs based on orientation-patterned gallium arsenide (OP-GaAs) crystals pumped by Cr:ZnSe or Tm:fiber femtosecond lasers have made it possible to reach wavelengths beyond ~4.8 µm, a barrier for the well-established oxide-based materials [7, 8]. Both DFG and OPO sources have been used for MIR optical frequency comb spectroscopy with different detection methods, namely a Fourier transform spectrometer [9-13], a virtually imaged phased array (VIPA) [14, 15], mode-resolved Vernier spectroscopy [16], and dual comb spectroscopy [17, 18]. However, the spectral range of all previous demonstrations was limited to <4.8 µm.

Here we report a versatile optical frequency comb spectroscopy system based on a doubly resonant optical parametric oscillator (DROPO) with an OP-GaAs crystal operating in the 3-5.4 µm wavelength range. The system incorporates two detection methods, a fast-scanning Fourier transform spectrometer (FTS) in combination with a multipass cell, and a continuous-filtering Vernier spectrometer. The FTS provides ultra-broadband spectral coverage, absolute frequency calibration and high precision in the absorption measurement. We demonstrate this by acquiring absorption spectra of CH$_4$, NO, and ambient air in the signal and idler ranges and comparing the results to the theoretical models. On the other hand, the Vernier spectrometer, which is intrinsically a cavity-enhanced system, is faster, more robust and compact. In the initial demonstration of mid-infrared Vernier spectroscopy we measure spectra of ambient air at the signal wavelengths.

The overview of the experimental setup is shown in Figure 1. The DROPO is synchronously pumped by a mode-locked Tm:fiber laser with a repetition rate of 125 MHz that delivers ~90 fs pulses with maximum 2.5 W average power in the 1.94-1.97 µm range. The repetition rate, $f_{rep}$, of the pump laser is locked to an RF source referenced to a Rb clock and the carrier-envelope offset frequency is free running. The DROPO is configured in a 2.4 m long ring cavity, comprising a dielectric coated input coupler (IC), two concave gold mirrors (M$_{1,2}$) with 50 mm radius of curvature, and three plane gold mirrors (M$_{3-5}$). The IC (Lohnstar, ZnSe substrate) has high reflectivity (R>99%) in the 3-6 µm range and it is anti-reflection coated for the pump wavelengths (T>85%). The Brewster-cut 0.5-mm thin OP-GaAs crystal has a fixed reversal period of 53 µm for quasi-phase-matching (QPM) at room temperature. A half-wave plate placed in front of the DROPO provides the pump polarization required for QPM. Mirror $M_5$ is attached to a piezo electric transducer (PZT) and mounted on a motorized translation stage (TS) to allow tuning of the DROPO cavity length. Up to 48 mW of combined signal/idler average power is coupled out using a thin film pellicle beam splitter (output coupler, OC, Thorlabs, BP145B4) deployed at Brewster angle inside the cavity. The residual pump from the OC is blocked using a long-pass filter (LPF, Edmund Optics). The signal/idler beam is aligned either to a multipass cell and a fast-scanning Fourier transform spectrometer (FTS) or to an enhancement cavity and a diffraction grating with a photodetector configured as a continuous-filtering Vernier spectrometer. The DROPO output is also monitored for stabilization and diagnostic purposes with a PbSe photodetector ($PD_1$, Thorlabs, PDA-20H-EC) placed after the IC. The DROPO is enclosed in a box and purged with nitrogen to minimize the absorption by atmospheric species inside the cavity.

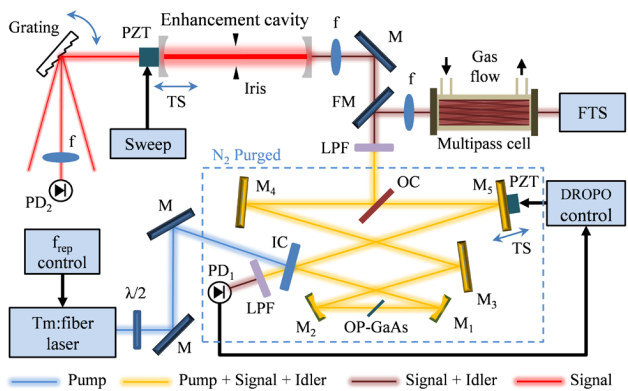

Fig. 1. Schematic of the experimental setup. λ/2: Half wave plate, M: Gold mirrors; IC: Input coupler; $M_{1-5}$: DROPO cavity mirrors; OC: Output coupler, PZT: Piezo electric transducer; TS: Translation stage; LPF: Long pass filter, FM: Flip mirror; f: Lens; $PD_{1-2}$: Photodetectors.

Doubly resonant operation of the DROPO (in degenerate and non-degenerate mode) is achieved at a discrete set of cavity lengths separated by the pump laser center wavelength [8]. A few typical DROPO output spectra recorded with the FTS are shown in Fig. 2. The absorption lines are due to the presence of atmospheric species (mainly $H_2O$ and $CO_2$) in the unpurged beam path between the DROPO and the detector in the FTS. No dispersion compensation component is used in the DROPO. Continuous operation at a particular signal/idler wavelength range is achieved by stabilizing the DROPO cavity length using a dither lock [19]. A low-frequency sine modulation is applied to the PZT and an error signal is obtained by synchronously demodulating the output of $PD_1$. The correction signal is fed back to the PZT via a proportional-integral controller.

The main advantage of the detection system based on the FTS with the multipass cell is that it allows spectroscopic measurements over the entire tuning range of the DROPO [9, 20]. Moreover, the spectra are free from any influence of the instrumental lineshape function [21]. The multipass cell (Aerodyne Research, AMAC-76LW) is ~32 cm long and configured for 238 passes, which results in 76 m optical path length, and it is connected to a gas flow system and a vacuum pump. A single spectrum with 250 MHz or 1 GHz resolution is acquired with the FTS in 1.5 s and 0.4 s, respectively. The optical path difference is calibrated with a stable cw diode laser at 1563.87(1) nm (Redfern Integrated Optics). The MIR interferogram is acquired with two HgCdTe detectors (VIGO, PVI-4TE-6) in balanced configuration connected to a high-resolution digitizer (National Instruments, PCI-5922) and a computer. To verify the performance of the spectrometer we measure absorption spectra of calibrated gas samples of $CH_4$ (4.9 ppm in 100 Torr of $N_2$) at 3.3 μm and NO (10 ppm in 100 Torr of $N_2$) at 5.2 μm. These spectra are shown in black in Fig. 3 together with fits of the corresponding model spectra calculated using absorption line parameters from the HITRAN database [22] and Voigt profiles with concentration as the fitting parameter (red, inverted). The spectra are normalized to background spectra measured when the multipass cell is filled with $N_2$, and a sum of a 3rd-order polynomial and low-frequency etalon fringes is used to correct the remaining baseline. Data points where atmospheric water absorbs nearly 100% in the path outside the multipass cell and prevents normalization are removed from the NO spectrum. The lower signal to noise ratio (SNR) in the $CH_4$ spectrum is caused by the higher attenuation by the multipass cell at the signal wavelength range. The residuals, shown below each panel, indicate that the general agreement with the model is very good. The remaining structure is presumably caused by inaccuracies in the HITRAN database and the use of the Voigt lineshape, and by a nonlinearity in the FTS detectors. The concentrations returned by the fits are 4.91(16) ppm for $CH_4$ and 9.80(23) ppm for NO, where the error is the standard deviation from ten consecutive measurements. Both results are within the accuracy limits of the corresponding gas samples (2%) and the gas flow and pressure meters (2% each). To demonstrate the ability of the system to discriminate between different species in a complex mixture we measure spectra of ambient air at atmospheric pressure. Figure 4 shows in black the normalized spectra of air (in the multipass cell) in the signal (a) and the idler (b) wavelength ranges. To retrieve the concentrations of the different atmospheric species we fit the sum of their model spectra based on the HITRAN database parameters (in color, inverted). Since many water lines (red) absorb nearly 100% we plot transmission instead of absorption and we do not take these lines into account in the fitting process. The signal wavelength range contains absorption features of 0.57(2)% of $H_2O$ and 1.94(6) ppm of $CH_4$ (blue), while the idler wavelength range reveals the presence of 469(29) ppm of $CO_2$ (purple) and 46(17) ppb of CO (green), in addition to 0.53(3)% of $H_2O$. The zoomed-in spectra and the residuals of the fits show the good agreement between the measurement and the model. The absorption sensitivity varies across the tuning range of the DROPO because of the different peak intensities and widths of the signal and idler spectra. For the 250 MHz resolution spectra shown in Fig. 3, the noise equivalent absorption (NEA), calculated as the standard deviation of the noise on the baseline normalized to the square root of the measurement time is $1 \times 10^{-5}$ $cm^{-1}$ $Hz^{-1/2}$ for the signal and $2.2 \times 10^{-6}$ $cm^{-1}$ $Hz^{-1/2}$ for the idler. Normalized to the number of spectral elements (~37000 for the signal, ~8700 for the idler), this yields a figure of merit of $5.2 \times 10^{-8}$ $cm^{-1}$ $Hz^{-1/2}$ per spectral element for the signal and $2.4 \times 10^{-8}$ $cm^{-1}$ $Hz^{-1/2}$ per spectral element for the idler wavelength range. Using the multiline fitting method [9, 23] we estimate the minimum detectable concentration to be 20 ppb $Hz^{-1/2}$ for $CH_4$, 15 ppb $Hz^{-1/2}$ for NO (both at 100 Torr), 10 ppb $Hz^{-1/2}$ for CO and 7 ppm $Hz^{-1/2}$ for $CO_2$ (both at 760 Torr).

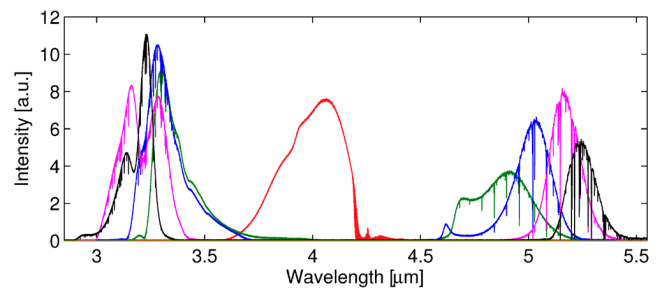

Fig. 2. The DROPO output spectra measured with the FTS for different cavity lengths and pump laser center wavelengths.

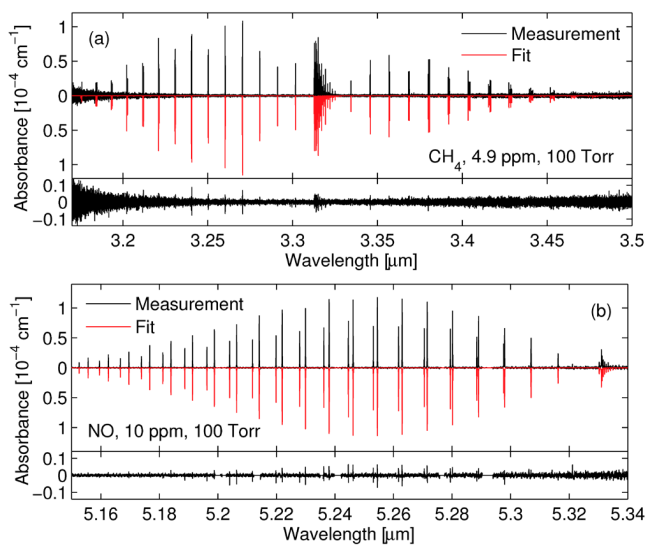

Fig. 3. Absorption spectra (black, no averaging, 250 MHz resolution) of (a) 4.9 ppm of $CH_4$ and (b) 10 ppm of NO, both in 100 Torr of $N_2$, along with fitted spectra calculated using the HITRAN database (red, inverted for clarity), and the residuals of the fit (lower panels).

For applications that do not require the absolute frequency calibration provided by the FTS, but instead need a faster and more compact detection system, we have implemented a continuous-filtering Vernier spectrometer [24, 25]. The system is based on an enhancement cavity, a diffraction grating and a photodetector. The cavity has a finesse ($F$) of ~300 at ~3.3 μm and is made of two highly reflective concave mirrors (Lohnstar, 5 m radius of curvature) on ZnSe substrates separated by $L = 60$ cm and open to air. An iris in the middle of the cavity is used to remove the remaining higher order transverse cavity modes. The output mirror is attached to a PZT and mounted on a translation stage for varying the cavity length. When the cavity free spectral range is precisely matched to twice the $f_{rep}$ of the DROPO (250 MHz), every other signal comb line is coupled into a cavity mode and transmitted through the cavity. When the cavity length is detuned from this perfect match position by $\Delta L = k\lambda_{OFC}$, where $k$ is an integer and $\lambda_{OFC}$ is a wavelength within the spectral range of the signal comb, the cavity resonances act as a filter and transmit groups of comb modes around $\lambda_{OFC}$, called Vernier orders. As $\Delta L$ increases, the consecutive Vernier orders get closer to each other in the frequency domain and transmit fewer comb modes. This results in higher spectral resolution but also lower optical power in each order, and thus lower SNR in the Vernier spectra. Moreover, it has been shown that the contrast of absorption lines in the Vernier spectra is higher for a negative length mismatch [25]. Considering this and the tradeoff between the resolution and SNR we choose the -30th Vernier order, corresponding to $\Delta L$ of -99 μm. This order yields a spectral resolution of 10 GHz, calculated as $c/F\Delta L$. For this $\Delta L$ several Vernier orders lie simultaneously within the spectral range of the signal comb. A reflection diffraction grating (Thorlabs, GR1325, 450 grooves/mm) is placed after the cavity to spatially separate the adjacent Vernier orders, and a lens is used to focus the selected order onto a DC-coupled HgCdTe detector (PD2, VIGO, PVI-4TE-6). By dithering the cavity length with the PZT the Vernier orders are swept across the signal comb, and a piece of the transmitted spectrum (determined by the aperture of the lens) is imaged on the detector. Consecutive wavelength ranges of the cavity transmission are imaged onto the detector by rotating the grating. The PZT is scanned at 20 Hz and only the central linear part of the sweep lasting 2 ms is used for recording the spectra. The entire signal range of 3-3.4 μm is recorded in 16 steps, and the resulting spectrum obtained by attaching these pieces is shown in Fig. 5 (black). Motorizing the grating sweep and implementing a tracking servo to keep the chosen Vernier order on the detector via feedback to the cavity PZT will allow continuous acquisition of the entire spectrum [24]. Since the cavity is open to the air no background spectrum can be retrieved for normalization. Thus, each individual piece is normalized to its maximum value and a low order polynomial is used for correcting the baseline. The red and blue curves, inverted for clarity, show model spectra of $H_2O$ and $CH_4$, respectively, calculated using the line parameters from the HITRAN database, the concentrations found from the measurement with the FTS (i.e. 0.57% of $H_2O$ and 1.9 ppm of $CH_4$), and the wavelength dependent cavity finesse calculated using the reflectivity curve of the mirrors. In the model the cavity-enhanced spectrum is convolved with a 10 GHz wide Lorentzian function to account for the resolution of the Vernier order [25]. The absolute wavelength scale of the measured spectrum is found by comparing the positions of the measured and calculated lines and using a 2nd order polynomial function to compensate for the nonlinearity caused by the grating rotation. The spectral features of both molecular species are clearly resolved, as shown in the zoomed-in spectra. The discrepancy between the measurement and the model is caused by the narrowing of the negative Vernier order due to strong molecular absorption, which is not taken into account in the simple convolution model. The full model, which is computationally more demanding [25], will be implemented in future work. The relative noise on the baseline is equal to $3.5 \times 10^{-3}$, which yields NEA of $6.2 \times 10^{-7}$ cm$^{-1}$ in 2 ms, where the effective cavity length is calculated as $FL/\pi$ [24]. This shows the potential of the Vernier spectrometer to exceed the sensitivity of the FTS-based system by orders of magnitude, since similar NEA is obtained with the FTS after a few seconds of averaging, even though the effective lengths of the multipass cell (76 m) and the low finesse cavity (57 m) are comparable.

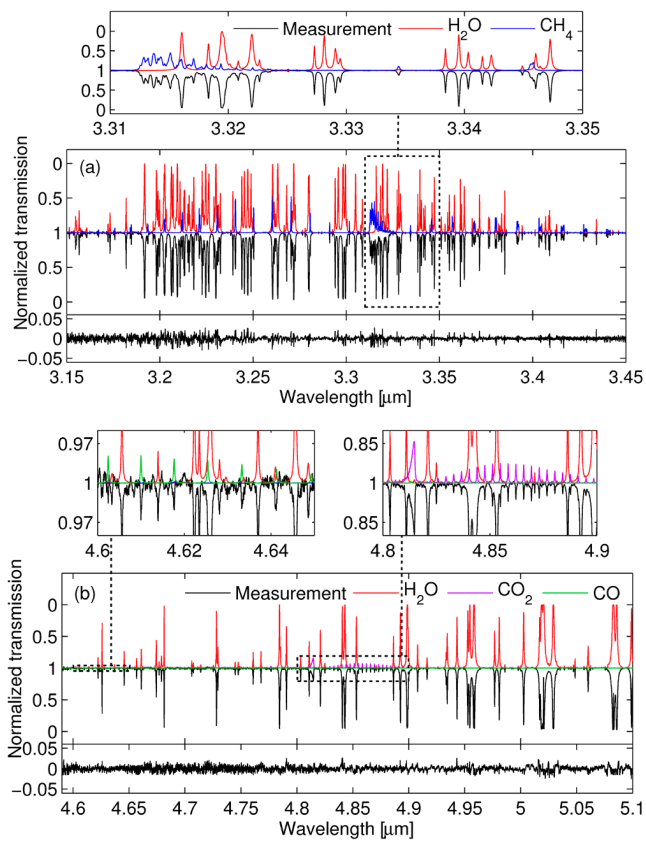

Fig. 4. Normalized transmission spectra (black, 10 averages, 1 GHz resolution) of air at atmospheric pressure in the (a) signal and the (b) idler wavelength ranges, along with the fitted spectra calculated using the HITRAN database for $H_2O$ (red), $CH_4$ (blue), CO (green), and $CO_2$ (purple), inverted for clarity, and the residuals (lower panels). The spectral ranges where the species other than water absorb are enlarged.

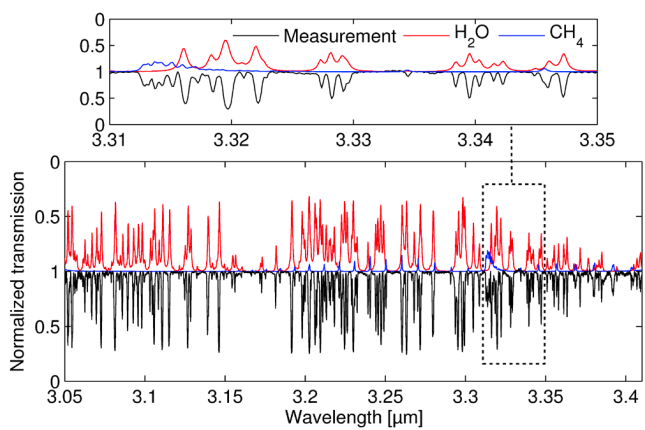

Fig. 5. Normalized transmission spectra (black) of atmospheric air measured by the Vernier spectrometer along with the model spectra of 0.57% of water (red) and 1.9 ppm of $CH_4$ (blue), inverted for clarity. The zoom shows the same spectral range as Fig. 4(a).

The two detection methods implemented in our system offer a number of complementary advantages. The FTS provides high resolution spectra with absolutely calibrated frequency axis over the entire tuning range of the DROPO. The gas concentration can be found with high accuracy by fitting a model to the spectra, and the lack of the instrumental lineshape function will enable precise measurements of absorption line parameters. However, a single spectrum is measured in acquisition times on the order of a second and the FTS is a rather bulky system. The Vernier spectrometer, on the other hand, is compact and robust, and allows the acquisition of a single spectrum in tens of ms. However, the spectral range is limited by the coating of the cavity mirrors, and only the signal (or the idler, but not both) can be recorded with one set of dielectric mirrors. The technique requires calibration of the frequency axis and the spectrum is convolved with an instrumental lineshape function. Using a higher finesse cavity the resolution of the continuous-filtering Vernier spectrometer can reach low GHz level, sufficient for measurements in atmospheric conditions. Moreover, the Vernier technique offers higher absorption sensitivity than an FTS-based system; should a cavity with the same finesse be implemented in both systems, at least an order of magnitude better sensitivity is expected with the Vernier scheme because of the much shorter acquisition time. In addition, using a cavity with the FTS requires a high bandwidth comb-cavity lock [10, 26], while a low bandwidth tracking servo is sufficient for the Vernier spectrometer.

In conclusion, we developed a mid-infrared optical frequency comb spectroscopy system that incorporates two complementary detection methods; an FTS in conjunction with a multipass cell for precision molecular spectroscopy, and a continuous-filtering Vernier spectrometer for faster and more robust measurements. The comb source, a femtosecond DROPO covering the 3-5.4 μm wavelength range, allows detection of a long list of molecules such as $CH_4$, NO, CO, $CO_2$, $H_2O$, as well as other hydrocarbons, $O_3$, HCN, $N_2O$, and $NO_2$. Of particular importance is the ability to address the fundamental N=O stretch transitions at 5.25 μm, which has not been demonstrated before with optical frequency comb spectroscopy to the best of our knowledge. The concentration detection limits for many molecules are in the low ppb levels, and both detection systems are capable of resolving their spectral features even in the presence of 100% water absorption. The continuous-filtering Vernier, demonstrated here for the first time in the MIR, offers a viable alternative to other cavity-enhanced systems based on FTS [10] or a VIPA [15].

**Funding.** The Swedish Research Council (621-2012-3650); Swedish Foundation for Strategic Research (ICA12-0031); Kempestiftelserna (JCK-1317.2); Carl Tryggers Stiftelse (CTS 14:145).

**Acknowledgement.** The authors thank Hsuan-Chen Chen for the initial work on the DROPO.